\documentclass[onecolumn,authoryear]{els-mrw}

\newcommand{\K}{\,\mathrm{K}}

\usepackage[tight,k-tight]{minitoc}

\usepackage{amsmath,amssymb,amsfonts,amsthm,makeidx,graphicx}
\usepackage{txfonts}
\usepackage{helvet}
\usepackage{cleveref}
\usepackage{subcaption,graphicx}
\mtcsetfeature{minitoc}{open}{\vspace{-0.3cm}}
\usepackage{tcolorbox}

\usepackage{ulem}
\usepackage{xcolor}

\begin{document}

\dominitoc
\chapter{Gas in Galaxies}
\label{chap:gas_in_galaxies}

\author[1]{Adam K. Leroy}
\author[2]{Alberto D. Bolatto}

\address[1]{Department of Astronomy, The Ohio State University, 140 West 18th Avenue, Columbus, OH 43210, USA}
\address[2]{Department of Astronomy and Joint Space-Science Institute, University of Maryland, 4296 Stadium Drive, College Park, MD 20742, USA}

\articletag{Chapter Article tagline: update of previous edition, reprint..}

\maketitle

\mtcsetfeature{minitoc}{after}{\vspace{-20pt}}
\mtcsetfeature{minitoc}{before}{\vspace{-20pt}}
\minitoc

\begin{glossary}[Glossary]
\term{Atomic gas} interstellar material composed mostly of hydrogen in its neutral atomic state. \\
\term{Molecular gas } interstellar material composed mostly of hydrogen in its molecular state.\\
\term{Ionized gas } interstellar material composed mostly of hydrogen in its ionized state. \\
\term{Main sequence of star-forming galaxies } well-defined relationship between star formation rate and stellar mass found for star-forming disk galaxies. \\
\end{glossary}

\begin{glossary}[Nomenclature]
\begin{tabular}{@{}lp{34pc}@{}}
ISM & Interstellar medium\\
{\sc Hi} & Atomic hydrogen \\
H$_2$ & Molecular hydrogen \\
{\sc Hii} & Ionized hydrogen \\
GMC & Giant molecular cloud\\
kpc & kiloparsec\\
\end{tabular}
\end{glossary}

\begin{abstract}[Abstract]
In this chapter, we give an overview of the major components of the interstellar medium (ISM) in galaxies at a level appropriate for upper level undergraduates or beginning graduate students. We discuss the major constituents of the the ISM in present-day star forming galaxies and summarize common methods to observe these components. We also review basic aspects of ISM structure accessible to extragalactic observations. Finally, we describe variations in ISM content and star-formation activity among local universe galaxies.
\\
\textbf{Keywords}: Extragalactic astronomy (506); Interstellar medium (847); Nebulae (1095); Disk galaxies (391)
\end{abstract}

\begin{tcolorbox}[
 standard jigsaw,
 colback=green!15,
 opacityframe=0.5,
 opacityback=0.2
]
\section*{Learning Objectives}
\begin{itemize}
  \item Understand the major components of the interstellar medium (ISM) in galaxies.
  \item Learn how astronomical observations assess the amount and properties of the ISM in galaxies via astronomical observations.
  \item Learn how the gas in galaxies is structured on scales accessible to observations of other galaxies.
  \item Understand how the abundance and properties of these ISM components vary from galaxy-to-galaxy. 
\end{itemize}

\end{tcolorbox}

\section{Introduction}
\label{sec:intro}

The gas and dust between stars in galaxies form the interstellar medium (ISM). This material represents the potential fuel for future star formation, and its contents and physical state play an important role in the growth and death of galaxies. The ISM is multiphase and dynamic in nature, with gas in distinct chemical states and at different temperatures and densities. New stars will form when the gas becomes cold and dense enough to collapse under the influence of gravity. In the long term, this star formation activity is fueled by the accretion of new ISM material from the surrounding circumgalactic medium (CGM) and the intergalactic media (IGM), the ``cosmic web'' which is the ultimate repository of new material for galaxy growth. Once inside the galaxy, gas moves under the influence of the galactic potential, its own self-gravity, and the impact from ``feedback.'' This feedback is energy, momentum, and mass injected by stars and accreting black holes. It heats and reshapes the ISM, while also potentially changing its phase and chemical makeup. Stars also create heavy elements, which are returned to the ISM via stellar winds and various stellar death scenarios including supernova explosions. A significant fraction of these heavy elements are depleted from the gas and form into a solid phase, referred to as dust.

The gas and dust in galaxies also exert a large influence on how we perceive the Universe. Dust on average reprocesses $\gtrsim 50\%$ of the optical and ultraviolet (UV) starlight into infrared (IR) emission. Both absorption and emission lines are also visible from X-ray to radio wavelengths. Some of these represent important channels for the gas to cool, and many of them carry critical information on the heavy element abundances and physical properties of the gas.

The goal of this chapter is to survey the basic contents of the ISM in galaxies, and to describe standard methods for observing this gaseous material and assessing its physical state. We focus on this in Section \ref{sec:contents}, which makes up the bulk of the chapter. Then in Section \ref{sec:struct} we discuss how this interstellar material is distributed within galaxies at the scales accessible to extragalactic observations. The actual ISM content of galaxies, including the abundance of cold gas and dust, varies from galaxy to galaxy. For example, the ISM in our Milky Way makes up about $10\%$ of the baryonic mass of the galaxy, while in the nearby Small Magellanic Cloud, the ISM makes up $\approx 50\%$ of the total material. These variations follow regular patterns and we describe some of the most important ones in Section \ref{sec:variations}. 

We aim this chapter at a beginning graduate or upper level undergraduate student. As such, we emphasize a first-order, often schematic picture and provide references to recent review articles and textbooks. There an interested reader can dive into these topics in more depth. Also note that this chapter focuses on the contents, structure, and patterns of ``low redshift'' galaxies, meaning those seen in the nearby, present day universe. One of the most exciting developments of the last two decades has been the emergence of a rich set of observations capturing the gas in galaxies out to Cosmic Dawn. The physics and basic components of the ISM in galaxies do remain the same across cosmic time, but the balance of components evolve. Unfortunately, this is too much to capture in a single article, and we refer interested readers to \citet[][]{TACCONI20REVIEW} as an excellent reference for the evolution of gas across cosmic time.

\section{Components of the interstellar medium in low redshift galaxies}
\label{sec:contents}

\subsection{Neutral atomic gas}
\label{sec:hi}

Most of the gas in most star-forming galaxies, including our own Milky Way, is atomic gas, often referred to as \textsc{Hi}. In this phase, the hydrogen exists in neutral, atomic form. Atomic gas in galaxies exhibits a wide range of temperatures, $T \approx 50$~K to $T >5,000$~K, and a correspondingly large range of densities, $n \sim 0.1{-}100$~cm$^{-3}$. \citet{KALBERLA09REVIEW} and \citet{MCCLUREGRIFFITHS23REVIEW} review the content and structure of the atomic gas in the Milky Way. Our Galaxy harbors $\approx 8 \times 10^{9}$~M$_\odot$ of this atomic gas, with a typical surface density $\Sigma_{\rm gas}=8$~M$_\odot$~pc$^{-2}$. The \textsc{Hi} makes up $\approx 10\%$ of the Milky Way's total baryonic mass of $\approx 10^{11}$~M$_\odot$, and a majority ($\approx 2/3$) of the total gas mass. In addition to contributing a large fraction of the mass, the \textsc{Hi} fills a large fraction of the volume in galaxies. Atomic gas is visible along every line of sight from the Solar System through the Milky Way galaxy.

The atomic gas exists largely in two phases. There is a cold, dense phase referred to as the cold neutral medium, or CNM, which has temperature $T \sim 50{-}100$~K, and hydrogen volume density $n \sim 10{-}30$~cm$^{-3}$. There is also a warm, diffuse phase, referred to as the warm neutral medium or WNM, which has higher temperature $T \sim 6,000-10,000$~K and lower density $n \sim 0.5$~cm$^{-3}$. This warm component of the atomic medium fills a substantial portion of the volume in the disks of galaxies, $\sim 50\%$ in the Solar Neighborhood, and is probably intermixed with a diffuse ionized phase (\S\ref{sec:hii}). Meanwhile, the cold, dense phase is concentrated into denser structures, and more associated with star formation and molecular material.

The two phases exist in approximate pressure equilibrium, meaning that the product of $n$ and $T$ will be the same for the WNM and CNM. They share a typical thermal pressure $P / k_B = n T \approx 3,500$~cm$^{-3}$~K in the Solar Neighborhood. This pressure varies across galaxies, and is lower in low density outer regions of galaxy disks or low mass galaxies. Gas with intermediate temperatures and densities between the CNM and WNM has been observed and is sometimes labeled the ``unstable neutral medium'' (UNM). This likely reflects material recently driven out of thermal equilibrium, e.g., by turbulence or stellar feedback, and is more abundant in lines of sight at high Galactic latitude.

\textbf{The 21-cm line:} \textsc{Hi} can be directly observed using radio telescopes in $\lambda = 21$-cm line emission \citep{VANDEHULST45HI,EWENPURCELL51HI}. This ``spin flip'' hyperfine transition at $\nu = 1.42040580$~MHz ($\lambda \approx 21$~cm) has a low Einstein $A$, and because of its low energy the condition $h \nu \ll k T$ is fulfilled for all reasonable ISM temperatures. As a result, the two available states are almost always populated according to their statistical weights, i.e., in 3-to-1 ratio. Because the level populations are effectively known, when the 21-cm line is optically thin its intensity can be translated directly to a column density, i.e., a number of H atoms per unit area. Similarly, a measured 21-cm flux from a source at a known distance can be translated into a total mass of \textsc{Hi} via $N_{\rm HI} = 1.823 \times 10^{18}~{\rm cm}^{-2} ~\left( \frac{\int T_{B,{\rm HI}} ~dv}{1 {\rm K~km~s^{-1}}} \right)$ with $\int T_{B,{\rm HI}} ~dv$ the line-integrated 21-cm intensity in brightness units \citep[e.g.,][]{DRAINE11BOOK,CONDON16BOOK}.

\begin{figure}
\centering
\includegraphics[width=0.8\linewidth]{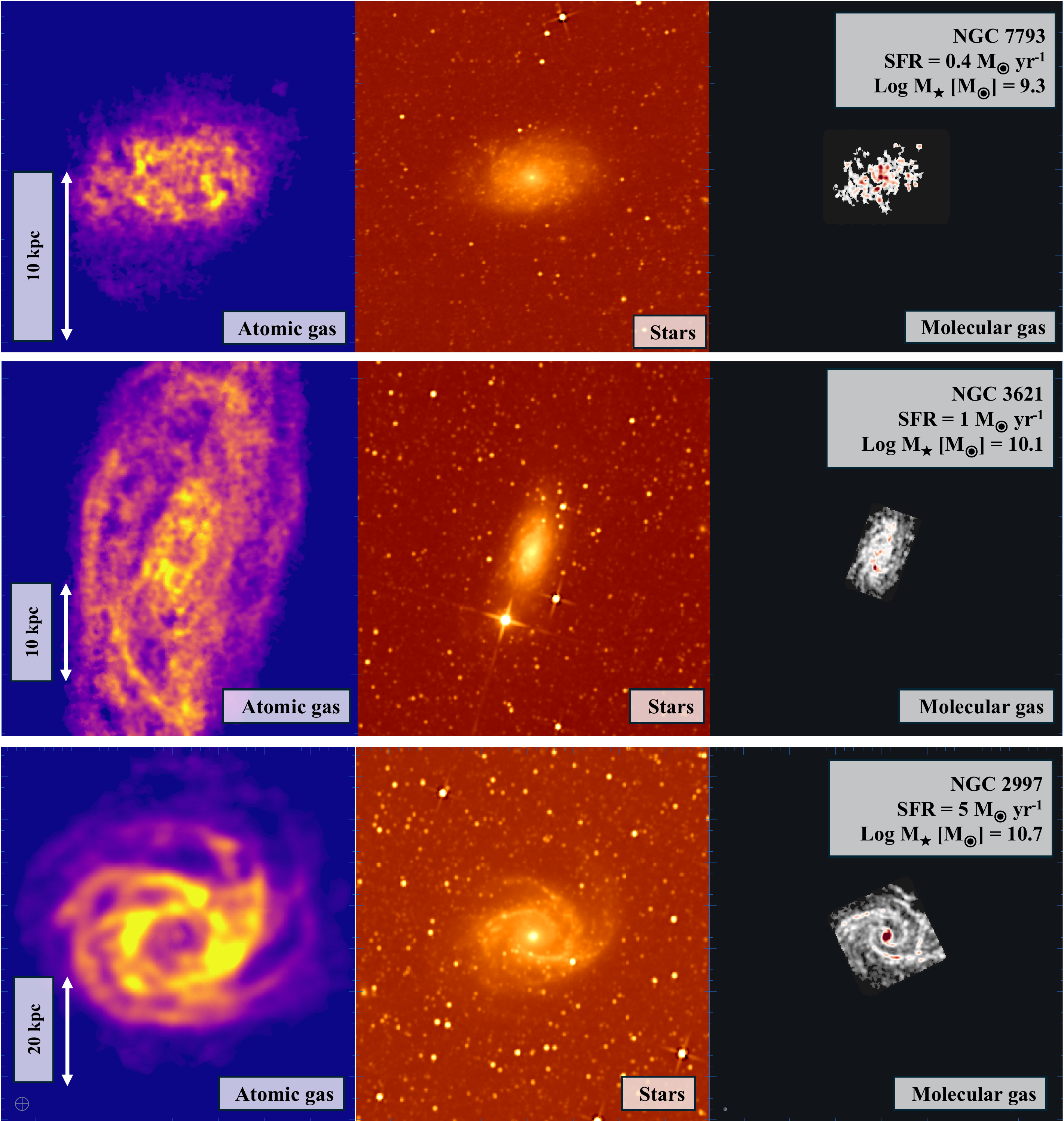}
\caption{Wide area view of the atomic gas traced by 21-cm observations (left),  near-infrared emission from stars (middle), and CO~(2-1) emission tracing molecular gas in three galaxies of different stellar mass. The figure demonstrates the extended atomic gas reservoirs often found in star-forming field galaxies, which can extend well beyond the stellar disk. The colder, denser molecular gas tends to be follow the distribution of the stars.
\label{fig:widefield}}
\end{figure}

The 21-cm line has been surveyed across the sky by increasingly powerful radio telescopes since the first extragalactic detections by \citet{KERR1954HI} (e.g., Fig. \ref{fig:widefield}). The large single dish Arecibo telescope conducted many of the current field-leading surveys of integrated \textsc{Hi} content from galaxies, while Very Large Array and Westerbork Synthesis Radio Telescopes (both radio interferometers) made major contributions to our knowledge of the structure of atomic gas in galaxies. In the present landscape, the new Five-hundred-meter Aperture Spherical Telescope (FAST), the MeerKAT and ASKAP arrays, and --- in the future --- the Square Kilometer Array and Next Generation Very Large Array are all powerful observatories where observing atomic gas via the 21-cm line represents an important part of the science mission.

\textbf{Other ways to trace the neutral atomic gas:} The UV Lyman $\alpha$ line (which is the $n=2\rightarrow1$ electronic transition of hydrogen) can be seen in absorption against background stars and quasars. Such observations have been critical to map out the extended structure of galaxies and the circumgalactic medium. Electronic transitions of other species with appropriate ionization potentials, for example sodium, are also used to trace the presence of atomic gas. Infrared fine structure lines, including the important \textsc{[Cii]} cooling line at $\lambda=157.7\mu$m, also emerge from the atomic phase of the ISM. And dust (see below) is mixed with atomic gas, so that the \textsc{Hi} can be traced by extinction or dust emission mapping. Because most of these techniques require challenging space-based observations, favorable geometries, and often-uncertain corrections for excitation, abundance, or other factors, they mostly play a complementary role to 21-cm studies.

\textbf{Separating the phases:} The 21-cm line offers a straightforward, reliable way to trace the mass of atomic gas, but does not on its own distinguish easily between cold, dense material (CNM) and warm, diffuse material (WNM). In the Milky Way, separating the CNM from the WNM is typically done by contrasting 21-cm absorption observations towards a bright background source with nearby emission. Combining emission and absorption constrains the optical depth of the 21-cm line, the column density of \textsc{Hi}, and the spin temperature of the gas \citep[see review in][]{MCCLUREGRIFFITHS23REVIEW}. In the Milky Way, such studies reveal that about 60\% of the total 21-cm emission is due to the WNM \citep{Heiles2003}. Finding appropriate bright background sources behind other galaxies is challenging, so observational constraints on the balance of phases beyond the Local Group remain weak. It is possible to use the \textsc{Hi} line width or morphology in well-resolved observations, or use the \textsc{[Cii]} far infrared emission (preferentially emitted by the CNM) to establish the contribution from each phase. This remains an important area for future work.

\subsection{Molecular gas}
\label{sec:h2}

\begin{figure}
\centering
\includegraphics[width=1.0\linewidth]{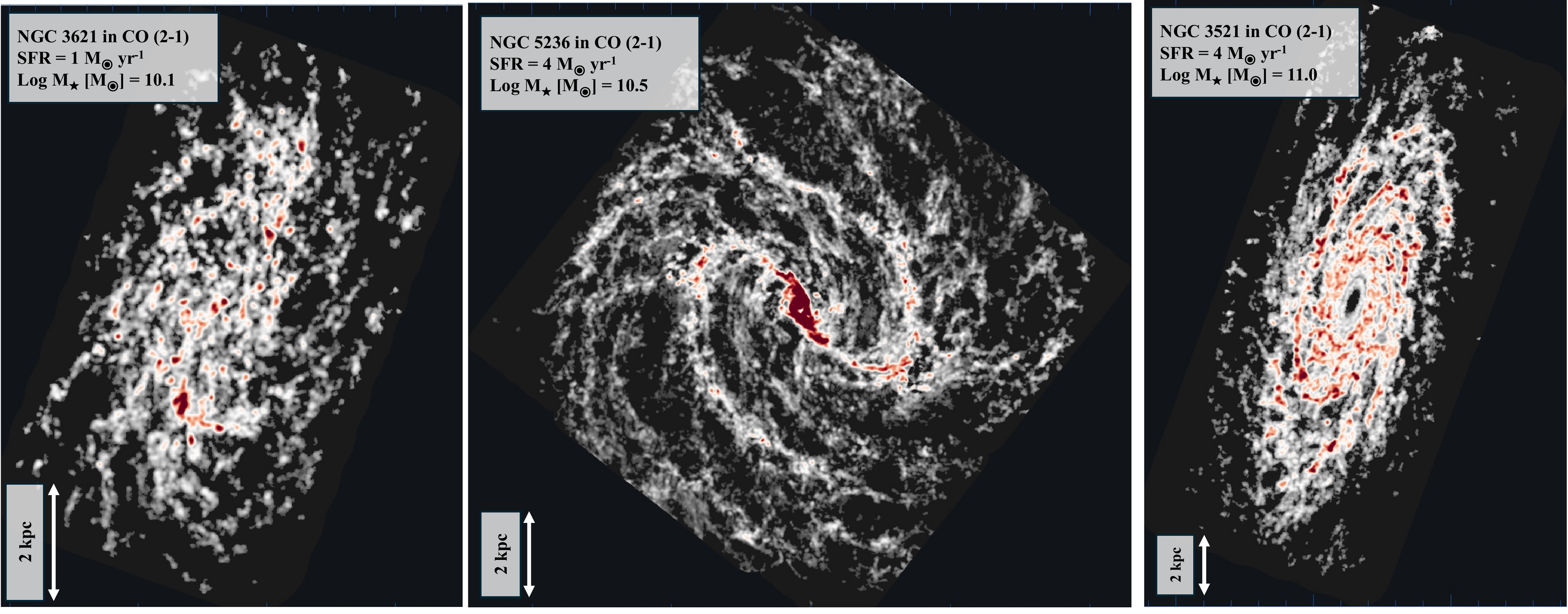}
\caption{CO~(2-1) maps tracing the molecular gas in three local universe galaxies (Section \ref{sec:h2}). The gas in the molecular phase is cold, mostly $T < 50$~K, shows a wide range of densities, $n_{\rm H2} \sim 10^{2}$~cm$^{-3}$ to $\gtrsim 10^5$~cm$^{-3}$. This gas is observed via rotational line emission of the second most common molecule, CO. The gas is concentrated mostly into giant molecular clouds, visible here as clumps in the CO map. The galaxies here show diverse large scale morphologies: Flocculent structure dominates NGC~3621 (left). In NGC 5236 (M83), strong spiral arms end at a stellar bar, which funnels material to a bright central molecular zone (middle). The massive NGC 3521 lacks molecular gas in the inner bulge and shows spiral structure at larger radii (right). Data from PHANGS--ALMA \citep{PHANGSALMA21SURVEY}.}
\label{fig:co}
\end{figure}

The molecular phase of the ISM consists of mostly cool ($T \approx 10{-}100$~K), dense ($n_{\rm H2} \approx 10^{1}{-}10^5$~cm$^{-3}$ or more) gas in which the dominant phase of hydrogen is molecular, H$_2$. In order for H$_2$ to become the dominant form of hydrogen, the gas must reach reasonably high densities and be shielded from dissociating UV radiation by a mixture of dust and self-shielding. This self-shielding occurs when dissociating photons are absorbed in discrete transitions,for H$_2$ the Lyman and Werner bands, that are optically thick because of the high abundance of the molecule; the high optical depth effectively reduces the photodissociation rate, protecting the molecule. As a result of these requirements, molecular gas is typically associated with high column densities, $N ({\rm H}) \gtrsim 10^{21}$~cm$^{-2}$, and line of sight dust shielding of $A_V \gtrsim 1$~mag \citep[see][]{WOLFIRE22REVIEW}. 

In contrast to the volume-filling atomic gas, the molecular gas mas is organized into dense structures that occupy a relatively small volume. These are often referred to as giant molecular clouds (GMCs). These structures are observed to be highly turbulent, with supersonic motions; their typical observed velocity dispersions are $3{-}10$~km~s$^{-1}$ compared to sound speeds $\ll 1$~km~s$^{-1}$). Each cloud also exhibits a wide range of densities and rich substructure \citep[e.g.,][]{HEYER15REVIEW,HACAR23FILAMENTS}. The densest substructures withing molecular clouds, with densities $n_{\rm H2} \gtrsim 10^5$~cm$^{-3}$, are often the ones observed to be directly associated with star formation. The mean properties of molecular clouds vary, but typical values are gas mass $M_{\rm gas} \gtrsim 10^5$~M$_\odot$, size $10{-}100$~pc, and surface density $\Sigma_{\rm gas} \gtrsim 100$~M$_\odot$~pc$^{-2}$ or $N({\rm H} \gtrsim 10^{22}$~cm$^{-2}$ \citep[see reviews in][]{FUKUI10REVIEW,HEYER15REVIEW,SCHINNERER24REVIEW}.

\textbf{Tracing molecular gas mass using CO line emission:} Because its lowest rotational transitions require $> 100$~K to excite, molecular hydrogen, H$_2$, is difficulty to observe directly at the temperatures found in molecular clouds. Fortunately, the CO molecule does emit readily under typical conditions found in molecular clouds. CO is the second most abundant molecule, typically $\sim 10^{-4}$ times as abundant compared to H$_2$, and represents dominant repository of gas-phase carbon within the molecular gas. The rotational transitions of CO are among the brightest emission lines from galaxies in the mm-wave part of the spectrum. The $J=1\rightarrow0$, $2\rightarrow1$, and $3\rightarrow2$ CO lines can be observed from the ground by telescopes including the ALMA, the IRAM facilities, the SMA, Nobeyama, and APEX. As a result, CO line emission has become the most common way to trace the distribution and kinematics of molecular gas in galaxies (e.g., Fig. \ref{fig:co}).

These CO lines can also be used to trace the molecular gas mass. Doing so relies on a theoretically and empirically calibrated CO-to-H$_2$ conversion factor, often referred to as $X_{\rm CO}$ or $\alpha_{\rm CO}$. This conversion factor is essentially a mass-to-light ratio that relates CO intensity to column density of H$_2$ and/or CO luminosity to total molecular gas mass. \citet{BOLATTO13REVIEW} review the physics of the CO-to-H$_2$ conversion factor in detail \citep[see also][]{Solomon2005,SCHINNERER24REVIEW}. The conversion factor is known to increase at low metallicities, where a ``CO-dark'' molecular gas phase becomes an important consideration. In such  phase, much of the gas is H$_2$ but CO is comparatively less abundant due to the lack of heavy elements and dust. Meanwhile the CO-to-H$_2$ conversion factor is lower in the centers of strongly barred spiral galaxies like the Milky Way and in merging galaxies. This reflects lower opacity in the CO, higher temperatures, and (for the higher $J$ CO lines) more excited CO molecules. Research on interpreting CO emission remains an active topic, and the precision with which molecular gas mass can be estimated from CO emission still limits many analyses.

\textbf{Other ways to trace molecular gas:} In addition to CO emission, dust emission and extinction or redenning have been widely used to trace molecular gas (see Section \ref{sec:dust}). Molecular hydrogen can be observed via UV absorption lines, but the need to observe UV background sources through high extinction molecular gas makes this technique challenging to apply to giant molecular clouds. Diffuse gamma ray emission, X-ray, and emission from other molecules, including OH or the subset of warm H$_2$ that emits in the infrared, have all also been used to trace molecular gas. Because the observations and/or analysis for many of these techniques remain challenging, they are still primarily used to calibrate the more readily observed CO and dust emission. See \citet{BOLATTO13REVIEW} for more.

\textbf{Physical conditions from molecular spectroscopy:} Molecular spectroscopy beyond only the $^{12}$CO lines offers powerful tools to constrain the excitation, temperature, density, opacity, and abundance of specific molecular species in the gas of other galaxies \citep[see reviews of techniques in][]{SHIRLEY15LINES,MANGUM15LINES}. These physical conditions are critical to understanding the nature and fate of the molecular gas. For example, the density of the gas sets the rate at which it will forms stars, while the temperature and opacity of the gas help determine the CO-to-H$_2$ conversion factor. Extragalactic observations can also access a number of relatively bright transitions associated with common molecules that have high Einstein A values and high critical densities. These include the low $J$ lines of HCN, HCO$^+$, CS, CN, all of which can be used to trace the balance between dense, excited gas and bulk gas. They are $>10$ times fainter than the $^{12}$CO rotational lines, and as a result the observations needed to infer density, excitation, or optical depth are more challenging. Even fainter transitions extending throughout the mm- and cm-wave part of the spectrum --- including transitions of N$_2$H$^+$, ammonia (NH$_3$), formaldehyde, SiO, and others --- provide a rich set of diagnostics, tracing temperature, density, the presence of shocks, etc. Many of these have been observed in individual sources, but their promise as general tools to studying in other galaxies mostly relies on future facilities like the Next Generation Very Large Array or an upgraded version of ALMA.

\subsection{Warm ionized gas}
\label{sec:hii}

\begin{figure}
\centering
\includegraphics[width=1.0\linewidth]{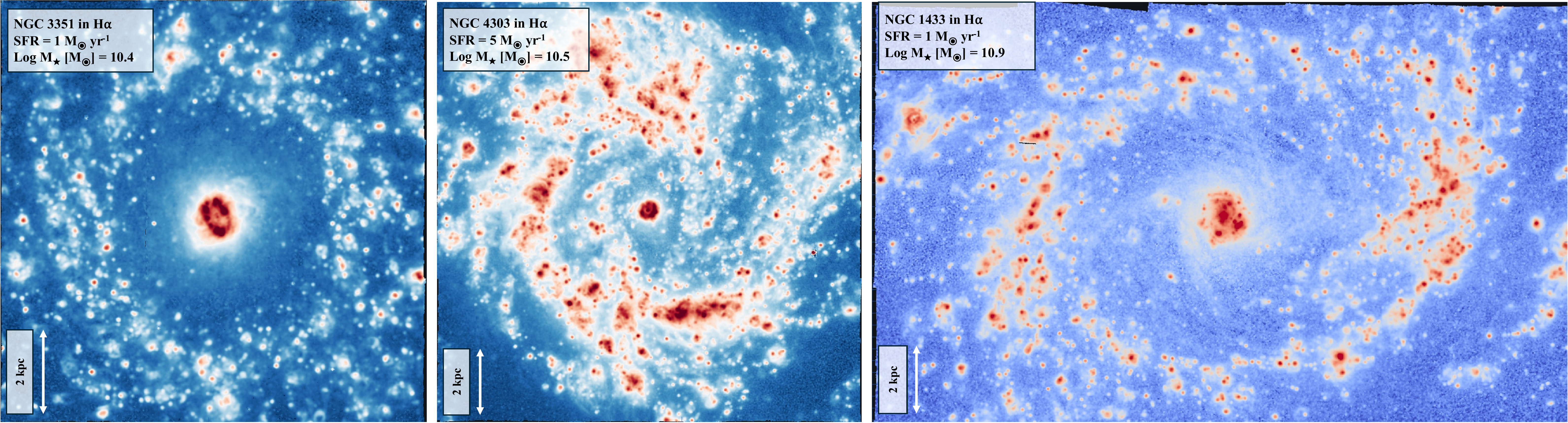}
\caption{H$\alpha$ line emission maps tracing the warm ionized gas and production of ionizing photons in three galaxies. The maps show H$\alpha$ recombination line emission, with the brightest clumps corresponding to individual \textsc{Hii} regions powered by massive, young stars. In between the \textsc{Hii} regions extended, fainter emission traces the diffuse ionized gas, also called the warm ionized medium (Section \ref{sec:hii}). In addition to revealing the distribution of ionized gas, the recombination line flux traces the production of ionizing photons, which in turn traces the mass of massive young stars and the rate of star formation (Section \ref{sec:formstars}). These H$\alpha$ maps are from the optical IFU survey PHANGS-MUSE \citep{PHANGSMUSE22SURVEY}. IFU surveys observe a rich suite of additional optical lines that can be paired to constrain the ionized gas density, temperature, extinction, ionization parameter, and powering source along each line of sight. Ionized gas associated with bar-fed central molecular zones are present at the center of each of these targets.
}
\label{fig:ha}
\end{figure}

Young, UV-bright stars, active galactic nuclei (AGN), and shocks all ionize interstellar hydrogen, creating a warm ($T \approx 5,000{-}15,000$) ionized medium where the dominant form of hydrogen is ionized \textsc{Hii}. Concentrations of this ionized gas surround young, massive stars as ``\textsc{Hii} regions,'' and these tend to be among the most visible ionized gas clouds in galaxies (see Figure \ref{fig:ha}).

Outside these visible nebulae, a lower density extended component of warm ionized gas pervades galaxies \citep[][]{HAFFNER09DIG,kewley2019,BELFIORE22DIG}. This extended ionized gas is referred to as the warm ionized medium (WIM) or diffuse ionized gas (DIG). The DIG contains much more mass than the \textsc{Hii} regions, but appears much fainter than \textsc{Hii} in line emission because the brightness of ionized gas emission lines is frequently proportional to the emission measure, which depends on the density of the gas squared, $EM \propto \int n^2 dl$. In the Milky Way, the DIG has mass $\sim 2 \times 10^9$~M$_\odot$, comparable to the mass of the molecular component. This gas is photoionized, and in star-forming galaxies the photoionization source for the DIG is likely dominated by ionizing radiation escaping \textsc{Hii} regions around young stars through pre-ionized channels. Other contributors are evolved stars and intermediate age binary systems, which are likely dominant in non-star-forming galaxies.

\textbf{Hydrogen recombination lines:} Ionized hydrogen recombines, with electrons becoming bound to protons. This produces bright line emission that traces the ionized gas in galaxies. The $n=3\rightarrow2$ Balmer $\alpha$ transition, H$\alpha$, is among the brightest optical emission lines from galaxies (e.g., Fig. \ref{fig:ha}). Other Balmer lines are visible through the optical, while the Paschen, Brackett, and fainter Humphreys and Pfund series are visible in the near-infrared \citep[][is the classic textbook on this topic]{OSTERBROCK06BOOK}. 

These recombination lines trace the location of ionized gas, but they do not trivially trace its mass because their intensity is proportional to the $EM\sim n^2$. However, because in equilibrium the rate of photoionization equals the rate of recombinations, optical and near-infrared recombination lines are excellent tracers of the ionizing photon production rate. Because most ionizing photons in star-forming galaxies originate from young, massive stars (when not in the presence of a nearby active galactic nucleus), these lines serve as critical tracers of the recent star formation rate (Section \ref{sec:formstars}). 

To do this, it is important to accurately account for the impact of dust extinction. Fortunately, given temperature and density of the gas the ratios among the hydrogen recombination lines at different wavelengths are known \citep[e.g.,][]{HUMMER87RECOMB,OSTERBROCK06BOOK}. Observations of recombination lines at widely spaced wavelengths, along with an assumed (or known) extinction curve, yields an estimate of the attenuation towards a \textsc{Hii} region. Combining many such lines can help constrain the extinction law and/or the source geometry \citep{CALZETTI94SFR,CALZETTI00DUST}.

\textbf{Physical diagnostics via emission line ratios:} Beyond hydrogen recombination lines, warm ionized gas produces a rich spectrum of UV, optical, and near-infrared emission lines. From ratios among the emission lines, e.g., of carbon, oxygen, sulfur, argon, and other abundant elements, one can infer the pressure, volume density of electrons, ionization parameter (a measure of radiation field strength divided by density), and temperature \citep[see review by][]{kewley2019}. 

Optical, UV, and IR line ratios also constrain metallicity, the abundance of heavy elements. Metallicity is often reported as $12 + \log_{10} {\rm O/H}$, with O/H the abundance of oxygen by number (the Solar abundance $12 + \log_{10} {\rm O/H} = 8.7$ is frequently also taken as the Solar Neighborhood abundance). The buildup of the elements over time is a crucial topic which has received significant attention, with a variety of methods deployed for different signal-to-noise regimes and types of galaxies \citep[see reviews in][]{kewley2019,SANCHEZ20REVIEW,MAIOLINO19REVIEW}. Measuring abundances precisely is challenging, and the exact results often depend on the methodology adopted \citep{KEWLEY08}. 

Ratios among emission lines from ionized nebulae can also be used to diagnose the nature of the source exciting the emission. BPT-type diagrams \citep{BALDWIN81,VEILLEUX87,CidFernandes2010} separate nebular emission likely to reflect conditions typical of \textsc{Hii} regions powered by young massive stars from those powered by shocks and harder radiation fields. This allows one to separate regions dominated by active galactic nuclei, but also large-scale shocks or even more extreme stellar populations \citep[][]{kewley2019,SANCHEZ20REVIEW}.

Optical emission lines can be observed from the ground and so have been studied for many years. Most major ground-based facilities have spectroscopic capabilities, but for many years studies of ionized gas were limited to spectra from individual slits or fibers. A major advance in the last $20$ years has been the deployment of wide area ``integral field units'' that create optical spectroscopic maps, often covering a large portion of the spectrum with many key emission lines for every location across large parts of galaxies. The ATLAS$^{\rm 3D}$, CALIFA, SAMI, and MaNGA surveys \citep{Cappellari2011,sanchez2012,Croom2012,bundy2015}, as well as the Multiobject Spectroscopic Explorer (MUSE) on the VLT have played prominent roles \citep{bacon2010}.

\subsection{Hot ionized gas}
\label{sec:hot}

Hot, low density gas also fills a large volume in the interstellar medium of galaxies, including the immediate region around the Sun. Though it contributes only a modest amount of the overall ISM mass, it fills a significant fraction of the volume and is important to the topology and evolution of the ISM. For example, the hot gas can be associated with pre-ionized channels through which ionizing radiation escapes \textsc{Hii} regions. In normal star forming galaxies, this hot gas is shock-ionized by stellar feedback, especially supernova explosions. The earliest stages of supernova feedback are visible as supernova remnants \citep[e.g.,][]{GREEN14MWREMNANTS}, which appear in the X-Ray, optical, and radio emission. These last only a relatively short time, $\lesssim 50,000$ yr, however, and so tend to be less numerous than \textsc{Hii} regions \citep{LONG22SNR,LI24SNR}.

The hot gas is visible in emission as soft $\sim 0.1{-}0.5$~keV X-Ray emission, corresponding to temperatures of one to a few million Kelvin ($0.1$~keV $\approx 1.16 \times 10^6$~K). Lower, but still high (few $10^5$~K), temperature gas is also visible in Milky Way absorption studies \citep[e.g.,][]{BOWEN08FUSE} but hard to access within the disks of other galaxies. The X-Ray emission from the hot ionized gas in normal galaxies is faint compared to the sensitivity of existing X-Ray facilities. As a result, only a handful of mostly very nearby galaxies have high quality published maps of diffuse X-Ray emission from the disk itself \citep[e.g.,][]{WANG21XRAY,LOPEZ23XRAY}, though studies of X-Ray visible halos and the CGM have been more common. When sufficient signal to noise can be achieved, X-ray spectroscopy can reveal the temperature, density, and column density of foreground absorbing material. This makes next generation X-Ray facilities with sensitivity to soft X-rays powerful probes of a less-studied phase of the ISM.

\subsection{Interstellar dust}
\label{sec:dust}

\begin{figure}
\centering
\includegraphics[width=1.0\linewidth]{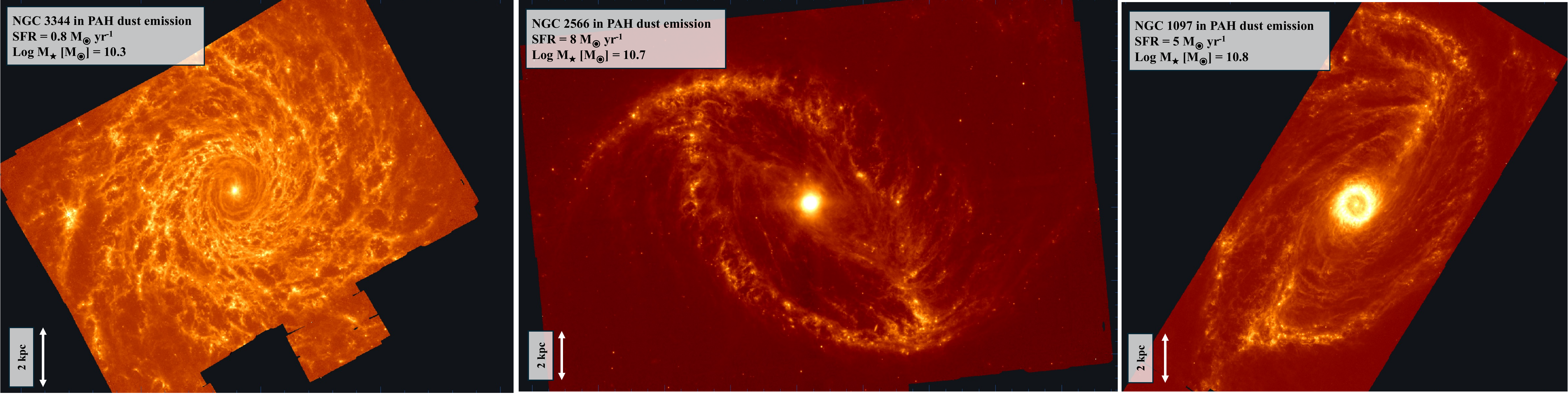}
\caption{Emission from polycyclic aromatic hydrocarbon (PAH) dust grains observed by JWST in three galaxies. Interstellar dust contains about half the metals and $\sim 1\%$ of the ISM mass in a galaxy like the Milky Way. It pervades galaxies, mixed with the gaseous phase of the ISM. The emission from dust reflects a mixture of the dust column density and the interstellar radiation field, which heats the dust. Here the same arms, flows along bars, and clouds visible in Fig. \ref{fig:co} are visible in the glow of dust grains. The PAH emission seen here reflects stretching and bending modes of PAH molecules, which produce broad features that dominate the mid-infrared spectrum of galaxies. Observations described in \citet{Chown2024}.}
\label{fig:pah}
\end{figure}

Solid grains of interstellar dust pervade galaxies alongside interstellar gas. In the Solar Neighborhood, this dust has mass $\sim 1\%$ the mass of gas. Refractory elements are heavily depleted from the gas phase into the dust, which encompasses as much as $40\%$ of the C and O and even larger fractions of Mg, Si, and Fe \citep[see][]{DRAINE11BOOK,GALLIANO18REVIEW}. Despite its modest mass, dust plays a crucial role in many aspects of gaseous ISM, including heating, shielding, and chemistry. It is also highly visible through the infrared and submillimeter parts of the spectrum and exerts a strong influence on the emergent optical and UV light from galaxies. \citet{GALLIANO18REVIEW} and \citet{HENSLEY21DUST} provide overviews of dust in nearby galaxies, including a synthesis of key observations of the Milky Way and Magellanic Clouds.

\textbf{Observing dust in emission:} Dust under typical interstellar conditions emits at infrared wavelengths, from $\sim 3{-}1000\mu$m. At typical ISM densities, $\lesssim 10^5$~cm$^{-3}$, the temperature of the larger interstellar dust grains is set by an equilibrium between radiative cooling and radiative heating by the interstellar radiation field (ISRF). An ISRF similar to that in the Solar Neighborhood leads to equilibrium dust temperature $T_{\rm dust} \approx 20$~K. This places the peak of the dust spectral energy distribution in the far infrared, $\lambda \approx 100{-}250\mu$m. Observing these wavelengths requires a telescope in space, or at least above most of the atmosphere. IRAS, then ISO, \textit{Spitzer}, \textit{Herschel}, SOFIA, and \textit{Akari} all made critical contributions to our understanding of far infrared emission from dust in galaxies, though there is no operational far infrared facility at this time.

Changes in intensity of the ISRF are often written as $U$, where $U=1$ corresponds to the Solar Neighborhood intensity. Taking into account the wavelength-dependent emissivity of grains leads to $T_{\rm dust} \propto U^{1/6}$. The ISRF is not directly observable, but the dust spectral energy distribution can be observed and $T_{\rm dust}$ inferred from observations at multiple wavelengths that span the peak of the SED. This makes estimating $T_{\rm dust}$ among the most practical methods of estimating the strength of the ISRF in the cold and neutral parts of the ISM.

Once the temperature is known, the optical depth of dust can be modeled. Physical dust models informed by observational constraints give the mass absorption coefficient, $\kappa$, which can be used to convert from the optical depth to the mass surface density of dust, $\Sigma_{\rm dust}$. This can be compared to measurements of the gas mass to determine the dust-to-gas ratio. Alternatively, if the dust-to-gas ratio is known (or can be estimated), $\Sigma_{\rm dust}$ can serve as an ISM tracer.

\textbf{Observing dust in absorption:} The properties and distribution of dust in galaxies can also be inferred from its effect on UV, optical, and even near-infrared light. Absorption and scattering by dust extinguishes light by a magnitude difference that depends on wavelength, $\Delta m=-2.5 \log(F_{obs}/F_{emit}) = -2.5 \log(e^{-\tau})$, where $\tau$ is the dust optical depth. This is the so-called ``general extinction'' $A_\lambda=1.086 \tau_\lambda$. The most common approach is to observe interstellar reddening (also called ``selective extinction''), which is the difference between two extinctions, $E(\lambda_1-\lambda_2)=A_{\lambda1}-A_{\lambda2}$. Extinction decreases as $\lambda$ increases, tending to zero when the wavelength is much larger than the dust grain size. Extinction as a function of wavelength (the ``extinction curve'') is characterized by the dimensionless parameter $R_V=A_V/E(B-V)$, where $B$ and $V$ are the usual optical photometric bands.  Extinction curves measured in the Milky Way can be parameterized so that $R_V$ describes them completely \citep{CARDELLI89DUST,Fitzpatrick1999,Gordon2023}. The $R_V$ ranges between 2 and 6 with larger values corresponding to denser and/or more extinguished regions, and a representative value is taken to be $R_v\approx3.1$ \citep{CARDELLI89DUST}. Dust and gas are intermixed, so that for a typical Milky Way dust-to-gas ratio at the Solar circle $N(H)=5.8\times10^{21} E(B-V)$\,H\,cm$^{-2}$\,mag$^{-1}$ where $N(H)$ represents the total column of hydrogen atoms in atomic and molecular form \citep{Bohlin1978,Rachford2009}. Measurements of the interstellar extinction curve extend to longer wavelengths \citep{Rieke1985,Rosenthal2000,Gao2013,Gordon2023}, showing a generally decreasing trend of extinction for increasingly long wavelengths but featuring a large bump at 10\,$\mu$m and another (less prominent) at 18\,$\mu$m, both due to silicates. There are variations in the extinction curve between galaxies, with the most notable variation the strength of the UV extinction bump at 2175\,\AA\ associated with small carbonaceous grains \citep[e.g.,][]{Gordon2003}. It is important to appreciate that, when determining extinctions over large areas containing ensembles of sources, both the amount and properties of the dust and the relative geometry of the dust and stellar distributions have an impact on the effective extinction:  a screen dust distribution in front of the sources produces different effects from an intermixed dust-stars distribution \citep{CALZETTI94SFR}.

\textbf{Polycyclic aromatic hydrocarbons, very small dust grains:} Polycyclic aromatic hydrocarbons (PAHs) are large molecules composed of several or many interlocked benzene-like carbon rings that on Earth are present in soot. PAHs may comprise up to 15\% of all ISM carbon \citep{Draine2003review,TIELENS08REVIEW,GALLIANO18REVIEW}. They are the smallest interstellar dust grains, and thought to be the carriers of the prominent mid-infrared spectral features observed at 3.3, 6.2, 7.7, 8.6, 11.3, 12.7, and 17 $\mu$m in galaxies (sometimes called aromatic features or aromatic infrared bands, AIBs). These features are broad ($\lambda/\Delta\lambda\sim3-10$) and bright, particularly in star-forming galaxies where they carry up to 20\% of the infrared luminosity. The different features are due to stretching and bending modes of C$-$H and C$-$C bonds. For example, the 3.3 $\mu$m feature (bright in small PAHs) is produced by C$-$H stretching, the C$-$C stretching gives rise to the 6.2 and 7.7 $\mu$m bands, in-plane bending of C$-$H causes the 8.6 $\mu$m feature, while many of the other features are associated with out-of-plane C$-$H bending. Ratios between PAH spectral features convey information on the size, structure, and charge of the molecules \citep{Draine2001,DRAINE11BOOK,Draine2021}. PAHs are somewhat fragile, and they can be destroyed in hot gas \citep{Micelotta2010hot} and shocks \citep{Micelotta2010shocks}, although they can also be produced in shocks by shattering of larger dust grains \citep{Jones1996}. They also appear to be destroyed in \textsc{Hii} regions produced by massive stars, as the emission from PAHs sharply decreases at the ionized gas boundary \citep{Compiegne2007,Peeters2024}. 

PAHs are at the lower end of a continuum of grain sizes which includes very small, small, and large dust grains \citep{galliano2018}. Very small grains are likely aggregates of PAHs and other carbonaceous materials \citep{Draine2003review}. Grains of all sizes, including PAHs, absorb non-ionizing UV and optical light which they re-emit primarily in the infrared \citep[although PAHs may also be responsible for emission in optical bands, see e.g.,][]{Witt2020}. Only large grains, however, are in equilibrium with the illuminating ISRF, reaching a well-defined stable temperature. Smaller grains, particularly grains under 50~\AA\ in size, experience large temperature spikes when they absorb a photon \citep{Draine2003review}. This phenomenon is called ``stochastic heating'' and causes emission by very small grains at short wavelengths, giving rise to a color temperature that is much higher than the blackbody equilibrium temperature \citep{Sellgren1984}. The combination of high sensitivity in the mid-infrared with high angular resolution and spectroscopy brought about by the \textit{James Webb Space Telescope} are opening a new window into our understanding of PAHs and their use to image and obtain new diagnostics of the ISM in galaxies (e.g., Fig. \ref{fig:pah}). 

\textbf{Dust as a tracer of gas and the dust-to-gas ratio:} Because dust is mixed with atomic, molecular, and (to some degree) ionized gas, mapping the dust column density also represents a powerful way to also map out the distribution of the ISM. This approach traces both atomic and molecular gas, including any CO-dark molecular gas and atomic gas where the 21-cm line is optically thick. This offers an advantage over 21-cm or CO mapping, and has enabled the use of dust as a tracer of otherwise invisible gas \citep[see][]{BOLATTO13REVIEW}. Converting from $\Sigma_{\rm dust}$ or $A_V$ to $\Sigma_{\rm gas}$ requires knowing the appropriate dust-to-gas ratio. In the Solar Neighborhood, depletion studies place this value at $\approx 1$-to-$150$ \citep[][]{DRAINE11BOOK,GALLIANO18REVIEW}. This dust-to-gas ratio varies from galaxy to galaxy and within galaxies as we discuss below (Section\ref{sec:abundance}).

\subsection{Forming stars}
\label{sec:formstars}

Star formation is closely linked to the ISM in galaxies, and recent or ongoing star formation is often traced by observations of the dust and gas \citep[see reviews in][]{KENNICUTT12REVIEW,CALZETTI13REVIEW}. As discussed above, recombination lines provide a powerful, direct probe of the ionizing photon production rate of the young stellar populations powering \textsc{Hii} regions. Correcting for the effects of interstellar extinction is critical, and can be done by observing multiple recombination lines at different wavelengths (see above). Radio free-free emission, which is robust to extinction, can also be used to directly trace the ionizing photon production rate in \textsc{Hii} regions \citep[e.g.,][]{MURPHY11SFR}. Current radio telescopes take significant time to observe the free-free emission from a typical star-forming galaxy, but the Next Generation Very Large Array should be highly efficient at mapping this extinction-free tracer of ionized gas.

Infrared emission is also widely used as a tracer of recent star formation. A large fraction of the dust UV and optical light produced by young stars is absorbed by dust and then re-radiated as infrared emission. This emission can be observed and used to infer the luminosity of the powering population. This approach has more ambiguity than recombination line studies, because the non-ionizing UV and optical emission that give rise to the IR are produced by stars that live over longer timescales than the massive stars producing the ionizing UV, and can be also generated by non-star forming stellar populations. On the other hand, the IR emission is sensitive to newly formed lower mass stars, and is readily available from all-sky imaging by the IRAS and WISE satellites \citep{WISE10} and extensive surveys by \textit{Spitzer}, \textit{Herschel}, and now JWST.

In addition to the ISM based methods, the recent (and not-so-recent) star formation rate can be inferred from modeling the stellar population. In more distant galaxies, population synthesis modeling of spectral energy distributions or spectra is the main tool \citep[see reviews in][]{CONROY13REVIEW,SANCHEZ20REVIEW}. In closer galaxies, individual stars can be observed and star formation histories reconstructed from their population statistics \citep[e.g.,][]{LEWIS15M31}. Focusing on the most recent star formation, direct counting of recently ($\lesssim 1$ Myr) formed young stellar objects, long considered a ``gold standard'' approach to measuring star formation in the Milky Way, is now possible throughout the Local Group of galaxies thanks to JWST \citep[e.g.,][]{peltonen2023}. Identification and analysis of star clusters represents an intermediate case between resolved stars and integrated light modeling and has also been a powerful tool \citep{KRUMHOLZ19REVIEW}. Finally, because UV light is mostly generated by young, massive stars, this light on its own has often been used as a tracer of recent star formation (in addition to being folded in to the more sophisticated analyses described above), although large corrections to correct for extinction by dust are necessary in dusty sources.

\subsection{Other components}
\label{sec:other}

Cosmic rays and magnetic fields are both energetically important but difficult to observe. Cosmic rays contribute significantly to the ISM pressure, are the dominant source of ionization in dust-enshrouded regions, and may plan an important role in driving galactic outflows or fountains \citep[e.g.,][]{THOMPSON24REVIEW}. Magnetic fields provide support against gravitational collapse and Galactic observations suggest that they have a role in shaping at least the atomic ISM \citep[e.g.,][]{MCCLUREGRIFFITHS23REVIEW}. Both of these phenomena are challenging to study in the Milky Way, and are even harder to access at intergalactic distances. Radio synchrotron emission, which is related to the magnetic field strength among other factors, offers an important probe, and is readily observed at low frequencies \citep[e.g.,][]{CONDON92REVIEW}. But this emission traces cosmic ray electrons, which are not the energetically dominant component. Magnetic field morphology is accessible through observations of infrared dust polarization and radio synchrotron polarization, but resolution is limited with current instruments and so only a limited number of systems can be studied.  Applying new diagnostics of cosmic ray flux and magnetic field strength, including those used in Milky Way studies, to other galaxies will be exciting directions for the next generation of infrared and radio telescopes.

\section{Structure of the gas in galaxies}
\label{sec:struct}

The ISM in galaxies is highly structured. The individual phases show structure on small scales, reflecting gravity, large-scale dynamical features and instabilities, and the balance of heating and cooling, molecule formation and destruction, and so on. On the largest scales, the gas often shows an extended radial distribution that represents the future reservoir for star formation and reflects accretion onto the galaxy by the circumgalactic and intergalactic medium. At intermediate scales, the gas responds to the galactic potential (which is often set by the stellar mass in the inner parts of galaxies) and to heating by stellar feedback. This sets both the vertical distribution of material and drives flows of gas through galaxy disks.

\subsection{Cloud-scale structure}
\label{sec:cloudscale}

Typically, high resolution extragalactic observations targeting nearby targets reach $\sim 50$~pc resolution. At this resolution, molecular gas appears clumpy and filamentary. Massive concentrations of molecular gas are called giant molecular clouds (GMCs). These appear to be the sites of most massive star formation. Though molecular clouds are not typically well-resolved at extragalactic distances, their sizes span the range $10{-}100$~pc with masses $\gtrsim 10^5$~M$_\odot$ of H$_2$, and they are known to harbor gas at a wide range of densities, $n_{\rm H2} \sim 10^2$ to $\gtrsim 10^5$~cm$^{-3}$. Their motions are dominated by supersonic turbulence, based on observed line widths that far exceed their thermal line widths. Their kinetic energy, assessed from spectroscopic CO imaging, approximately matches their gravitational potential energy so that they appear at least marginally bound \citep[][]{HEYER15REVIEW,SCHINNERER24REVIEW}. They are frequently viewed as existing in virial equilibrium, though observations also appear consistent with a fast evolving, even collapsing molecular medium \citep[][]{BALLESTEROS11GMCS}.

Emission lines from ionized gas resolve into discrete \textsc{Hii} regions at this resolution, as well as a smaller number of supernova remnants and planetary nebulae. These bubbles of ionized gas have sizes $\sim 5{-}15$~pc in Milky Way observations \citep{ANDERSON14SFR}, but are often described using a size-luminosity relation \citep[e.g.,][]{WISNIOSKI12HII} that extends to much larger sizes $\sim 100$~pc in extragalactic observations. They are associated with individual young stellar populations (though not always gravitationally bound clusters), which produce copious amounts of ionizing photons for the first $\sim 4{-}6$~Myr of their lifespan. 

Compared to the ionized and molecular gas, the atomic gas appears smooth, though it is worth emphasizing that 21-cm observations typically have significantly worse physical resolution than optical emission line or mm-wave CO observations. Still, observations of Local Group galaxies confirm this idea. \textsc{Hi} shows a relatively narrow range of column densities and fills a large fraction of the area in galaxies. The distribution is not perfectly smooth, however.  Large-scale shells and holes are often visible in the atomic gas \citep{bagetakos2011A,POKHREL20HOLES} as are dynamical features, including spiral arms. Some of these shells are carved by feedback, though others may naturally result from the flow of gas in the galaxy potential. The drivers of the \textsc{Hi} morphology and the density and temperature of the gas filling the \textsc{Hi} holes remain active topics of research.

Dust appears mixed with both atomic and molecular gas, and so the column density of dust mirrors the morphology of those components. Because radiation fields are more intense near young stars, the emission from dust near \textsc{Hii} regions is enhanced.

At $\sim 100$~pc resolution these different ISM phases are distinct. Their distributions are correlated on large scales, but not every molecular cloud appears coincident with an \textsc{Hii} region, and vice versa \citep{SCHRUBA10SFGAS,MURRAY11LIFETIMES}. A common explanation for this decorrelation is that the ISM clouds and star-forming regions are in different evolutionary states. Statistical modeling has been used to infer an evolution sequence for star-forming regions \citep{CHEVANCE20TIMES,KIM22TIMES}, finding that they evolve quickly likely due to the early impact of stellar feedback \citep[][]{chevance2023,SCHINNERER24REVIEW}.

\subsection{Radial structure}
\label{sec:radial}

\begin{figure}
\centering
\includegraphics[width=0.8\linewidth]{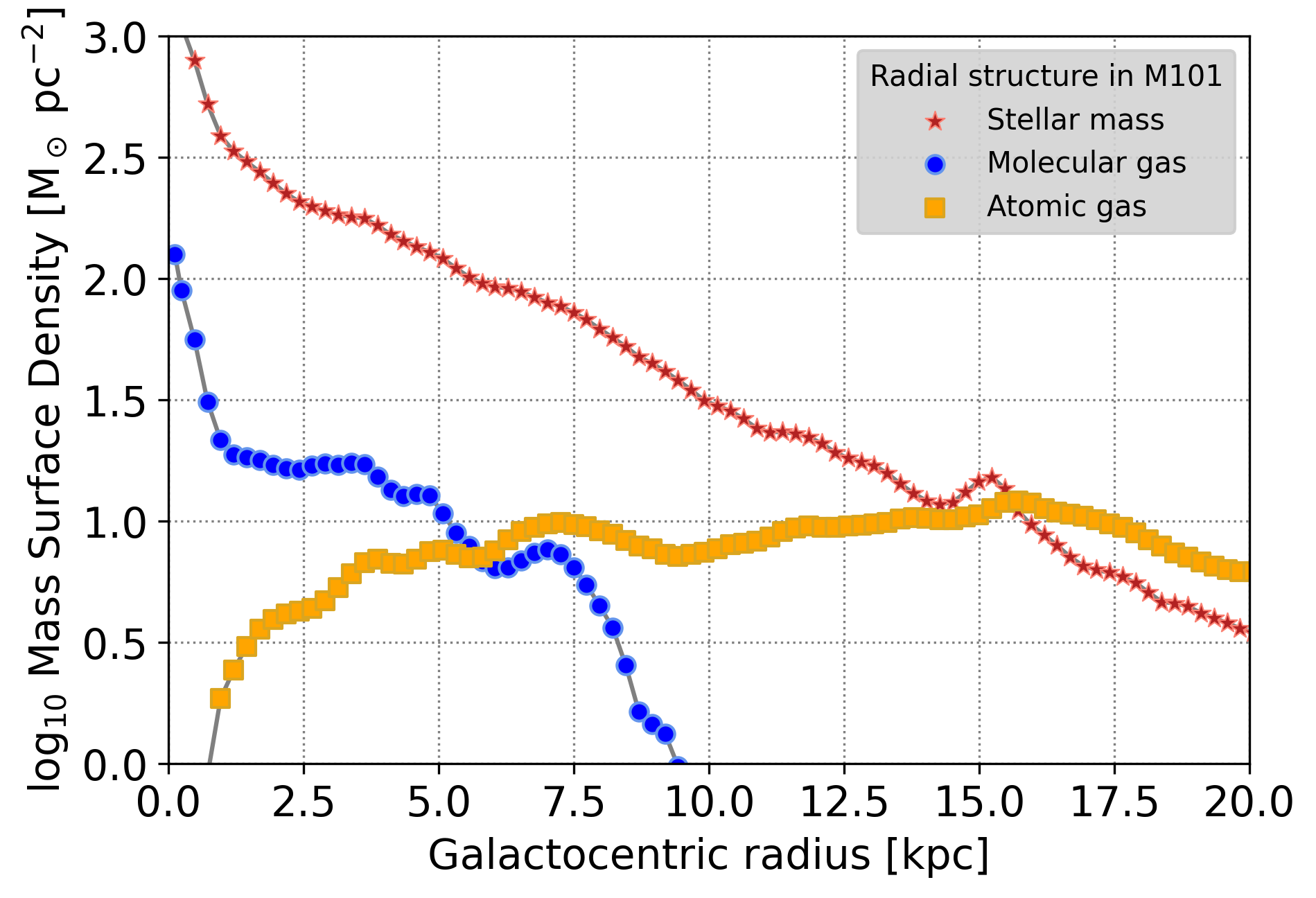}
\caption{Example of the radial distribution of mass surface density, measured in solar masses per square parsec, in the massive star-forming spiral galaxy M101. As in Fig. \ref{fig:widefield} the atomic gas shows an extended distribution with fairly constant mass surface density, while the stellar mass and molecular gas (estimated from CO) fall off exponentially on a shorter length scale. The stellar mass dominates over the ISM in the inner disk, but this situation changes at large galactocentric radii.
\label{fig:radial}}
\end{figure}

The ISM in field star-forming galaxies, as traced by \textsc{Hi} observations, often extends to very large galactocentric radius \citep{WANG16HI}. The extended atomic gas reservoir in the outer parts of galaxies, which often dominates the galaxy gas mass, frequently has higher surface density than the stellar disk \citep[][]{BIGIEL10SFGAS,WANG16HI}. The importance of the outer disk \textsc{Hi} likely reflects a combination of recently accreted material from the circumgalactic medium \citet{SANCISI08REVIEW}, and the inability of these extended reservoirs to convert most of their low density atomic gas into the cold, dense gas needed to form stars. Galaxies in dense large scale environments, such as clusters, often show extended \textsc{Hi} disks that are truncated or reshaped by interaction with intracluster gas or other galaxies \citep{CORTESE21REVIEW}. Therefore large scale environment has a significant impact on the atomic gas and overall ISM content of galaxies.

The molecular gas, stellar mass, and star formation activity all tend to follow a more compact distribution than the atomic gas. As a general rule, these three components track one another and follow an approximately exponential profile in star-forming disk galaxies \citep[e.g., Fig. \ref{fig:radial}][]{YOUNG91SFGAS,YOUNG95SFGAS,LEROY08SFGAS,BOLATTO17SFGAS}, though there are important deviations from this trend in both the inner and outer regions of galaxies. The more compact distribution of the molecular gas compared to atomic gas is understood to reflect the high surface density, high gas density, and high interstellar pressure in the inner parts of galaxies, which leads to a higher equilibrium ${\rm H}_2$/\textsc{Hi} ratio \citep{WONG02SFGAS,BLITZ06PRESS,LEROY08SFGAS}.

\subsection{Vertical structure and dynamical equilibrium}
\label{sec:vertical}

Unlike stars, gas is dissipative due to radiative cooling. Therefore, it settles in configurations that follow the potential of the galaxy. In gas-rich spirals, the ISM forms a thin disk, much thinner than the stellar component. In the Milky Way, the vertical distribution of the \textsc{Hi} component has an average FWHM of approximately 300~pc, although a better description in the inner Galaxy is two Gaussians with FWHM of 212 and 530~pc \citep[see][and references therein]{MCCLUREGRIFFITHS23REVIEW}. The thickness flares in the outer galaxy \citep{Levine2006}, following the lower vertical stellar potential. The molecular gas is concentrated on an even thinner layer with FWHM of $\approx 100$~pc within the Solar circle \citep{HEYER15REVIEW}.

The gas in galaxies self regulates to a vertical dynamical equilibrium, in which the weight of the gas is balanced against momentum and energy injected mostly by stellar feedback \citep{OSTRIKER10PRESS,OSTRIKER22PRESS}. Because the kinetic energy in the gas will decay on relatively short timescales, the gas will collapse in the absence of such feedback. When the gas begins to collapse it will compress, becoming denser and leading more gas to become molecular and star-forming. This leads to a higher star formation rate, more feedback, and expands the gas in the vertical direction. This expansion lowers the gas density, and therefore also the star formation rate. This negative feedback loop leads to self-regulation that describes the abundance of molecular gas and rate of star formation in galaxies well. The weight of the gas, often expressed as the pressure needed to maintain dynamical equilibrium, can be calculated from observables accessible in face-on galaxies \citep{OSTRIKER22PRESS,SCHINNERER24REVIEW}. Although we emphasize the vertical direction, similar self-regulation has been discussed in the radial direction, often using the Toomre's $Q$ formalism \citep[e.g.,][]{SILK97SFGAS,ELMEGREEN97SFGAS}.

\subsection{Stellar bars, arms, and nuclear starbursts}
\label{sec:bars}

Gas also responds to modes in the galactic potential, including spiral arms and stellar bars. Arms concentrate gas into spiral arms, which show much larger fractional enhancements in molecular gas surface density compared to stellar surface density \citep[e.g.,][]{MEIDT21ARMS}. Debates persist about whether the spirals arms only concentrate gas or also trigger molecular cloud and star formation via shocks or cloud collisions \citep[e.g.,][]{QUEREJETA24SFGAS}. In either case, they dominate the visible morphology of many disk galaxies.

Stellar bars also exert a large impact on the gas. Bars are present in $\approx 2/3$ of massive disk galaxies, including our own Milky Way. They drive significant radial gas flows along the bar towards the galaxy center. These flows are visible as dust lanes in optical and UV images and can be seen in emission in CO maps \citep[e.g.,][]{stuber2023}.

Fed by these bar-driven gas flows, molecular gas piles up in the center of many massive disk galaxies. These central molecular zones (CMZs) represent the most common ``starburst'' environment in the $z=0$ Universe \citep{KORMENDY04REVIEW,SCHINNERER24REVIEW}. They host large amounts of dense molecular gas and high concentrations of star formation activity. These CMZs represent the most prominent sites of super star cluster formation \citep[super star clusters are very massive and compact young star clusters caused by powerful localized star formation events,][]{PORTEGIESZWART10REVIEW} and frequently are the launching sites for powerful galactic winds \citep[][]{veilleux2020,THOMPSON24REVIEW}. Merging or strongly interacting galaxies achieve similar or even more extreme conditions, but are rarer in the present-day universe \citep{SANDERS96REVIEW}.

\section{Variations in gas content among galaxies}
\label{sec:variations}

\begin{figure}
\centering
\includegraphics[width=0.4\linewidth]{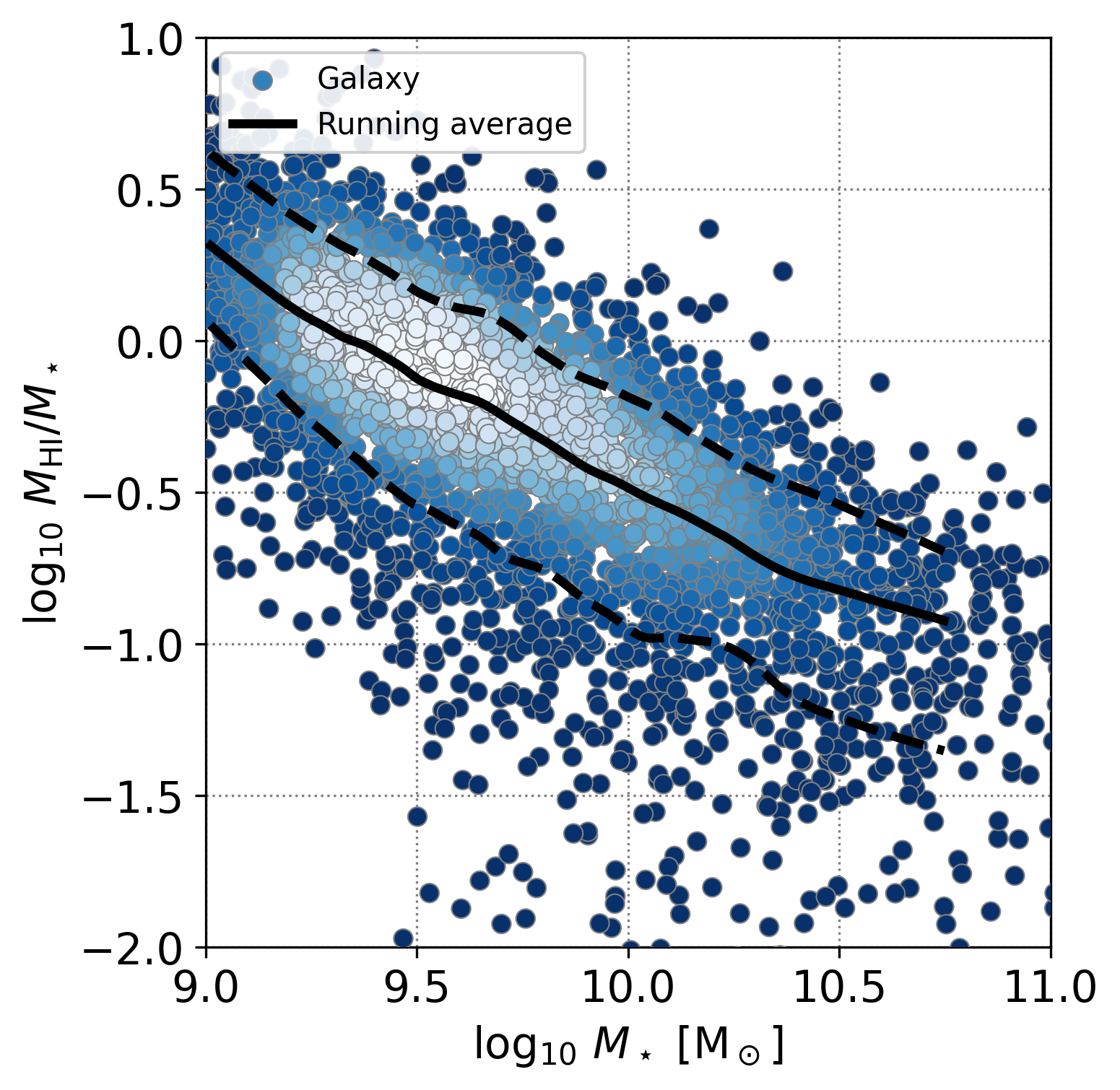}
\includegraphics[width=0.4\linewidth]{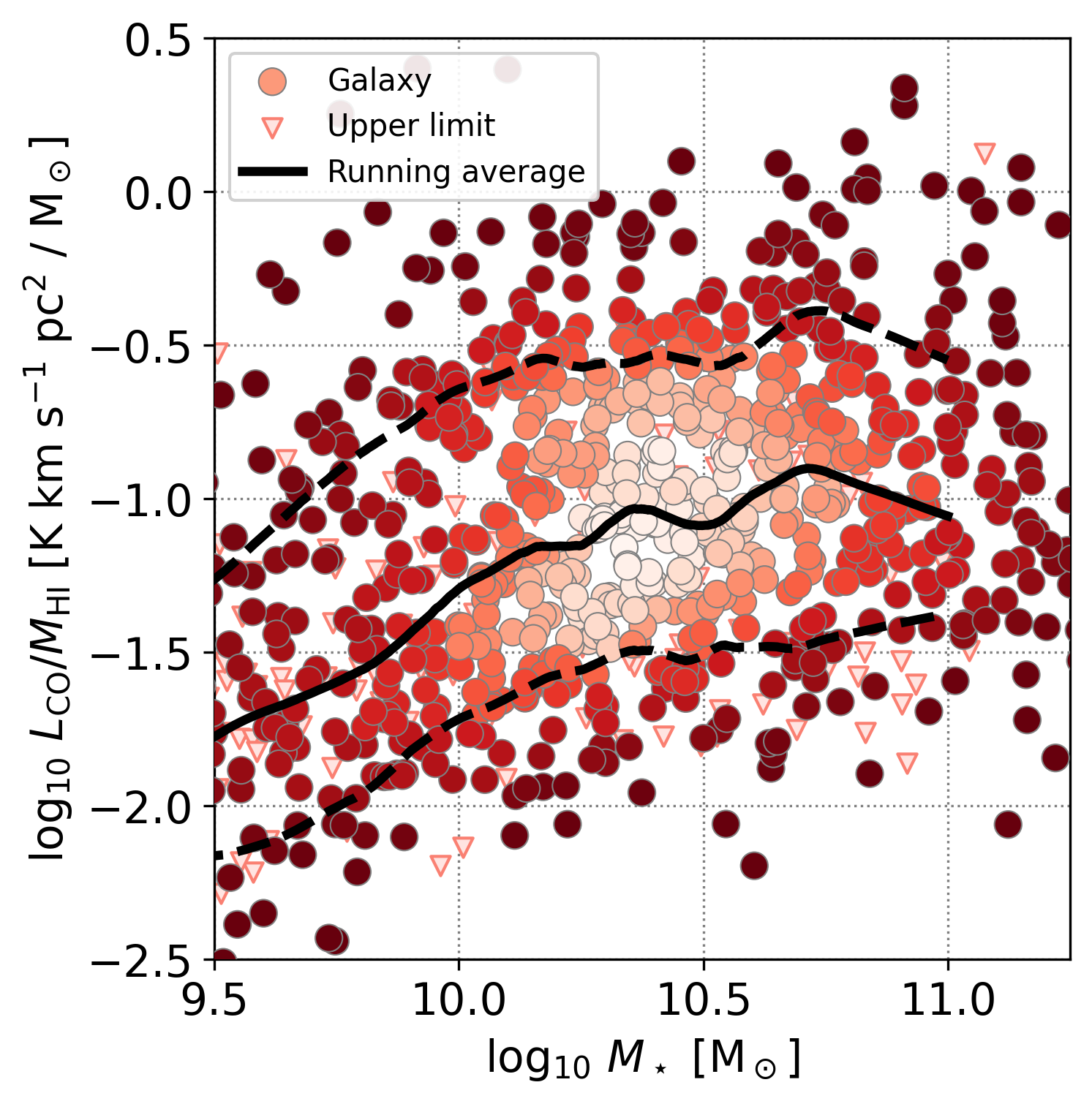}
\caption{Illustration of varying ISM content by galaxy mass. Each point here represents one galaxy with an integrated mass measurement from a large compilation drawn heavily from \citet{SAINTONGE17SFGAS} and \citet{Colombo2020} and \citet{Makarov2014}. The left panel shows that for star forming galaxies the ratio of atomic gas mass to stellar mass decreases with increasing stellar mass. The right panel shows how the ratio of CO luminosity, tracing molecular gas mass, to atomic gas increases with stellar mass. More massive star-forming galaxies have less fractional mass in their interstellar medium and more molecular compared to atomic gas. Color in the points reflects the density of data in that part of the plot. Upper limits in the right panel show galaxies where CO is not detected.
\label{fig:intscaling}}
\end{figure}

In Sections \ref{sec:contents} and \ref{sec:struct} we describe a general picture applicable to massive, star-forming $z=0$ disk galaxies. However, the gas content of galaxies, including the mixture of phases, varies. Much of this variation follows regular patterns. \citet{SAINTONGE22REVIEW} provide an excellent review of many of these variations and \citet{TACCONI20REVIEW} review how they change as a function of redshift. \citet{KENNICUTT12REVIEW} review the relationship between gas and star formation in massive $z=0$ disk galaxies, and \citet{SANCHEZ20REVIEW} review aspects of structure and scaling relations for $z=0$ galaxies based on IFU surveys.

\subsection{Star formation activity, stellar mass, and gas mass}
\label{sec:ks}

For star-forming disk galaxies, the star formation rate (SFR) of the galaxy correlates with the stellar mass, $M_\star$. This appears to hold true across redshift \citep[e.g.,][]{TACCONI20REVIEW}, and the relatively tight (rms scatter $\approx 0.3{-}0.4$~dex) relation is referred to as the ``main sequence of star forming galaxies.'' ``Starburst'' galaxies are often defined by having high SFR relative to galaxies with similar stellar mass, and so they reside above the star forming main sequence. Early type galaxies, ``green valley'' galaxies, or ``quenching'' galaxies have low SFR relative to that expected for their stellar mass (though note that all of these terms are also laden with additional meaning).

The star formation rate of a galaxy also correlates with the cold gas mass. The ratio of the two quantities expresses the ``specific'' star formation rate per unit gas, which is sometimes referred to as the star formation efficiency of the gas, ${\rm SFR}/M_{\rm gas}$ \citep{YOUNG91SFGAS,YOUNG96SFGAS}. Alternatively this quantity is expressed as the gas depletion time, $M_{\rm gas}/{\rm SFR}$, which expresses the time needed for current star formation to consume the entire gas reservoir. Specifically, the star formation rate of a galaxy tracks the molecular gas mass, as traced by CO \citep{YOUNG91SFGAS,YOUNG95SFGAS,KENNICUTT12REVIEW,SAINTONGE22REVIEW}, with typical molecular gas depletion times of $\approx 1{-}2$~Gyr for galaxies on the star forming main sequence at $z=0$ \citep{LEROY13SFGAS,TACCONI20REVIEW,SAINTONGE22REVIEW}. On galaxy scales, the star formation rate also correlates with the atomic gas mass, but the depletion time of molecular gas varies less than the atomic gas depletion time, and in studies that resolve galaxy disks, star formation activity correlates with molecular, rather than atomic gas \citep[][]{BIGIEL08SFGAS,LEROY08SFGAS,SCHRUBA11SFGAS}. Other galaxies thus apparently follow the same situation seen in the Milky Way, where stars form from clouds made of predominantly molecular gas. Because atomic gas represents more of the overall ISM mass in galaxies, this implies that the abundance of molecular gas can represent an important regulating factor in setting the star formation rate.

The Kennicutt Schmidt relation offers a popular alternative framework (instead of SFE or $\tau_{\rm dep}$) to express the relationship between gas and star formation. This approach \citep[][]{KENNICUTT89SFGAS,KENNICUTT98SFGAS} considers the surface densities of gas and star formation rate, $\Sigma_{\rm gas}$ and $\Sigma_{\rm SFR}$, as the primary variables and posits a power law relationship, $\Sigma_{\rm SFR} \propto \Sigma_{\rm gas}^n$ between them to describe galaxies. Considering atomic and molecular gas combined, the relationship has index $n \approx 1.5{-}2$ for integrated galaxies, depending on the sample studied and treatment of $\alpha_{\rm CO}$. Reflecting the more direct relationship between molecular gas and star formation, many studies focusing on the molecular gas Kennicutt Schmidt relation. These tend to find $n \approx 0.8{-}1.3$ \citep{BIGIEL08SFGAS,KENNICUTT12REVIEW,LEROY13SFGAS,SUN23SFGAS}.

The molecular gas depletion time varies \citep[see reviews in][]{SAINTONGE22REVIEW,TACCONI20REVIEW}. It appears shorter, indicating more efficient star formation, in the dense concentrations of gas found in bar-fed central molecular zones and merging galaxies. It also appears longer in quiescent galaxies, i.e., galaxies that have star formation rates lower than those predicted by the star-forming main sequence. It may also be shorter in low mass galaxies, but this result depends sensitively on still-uncertain knowledge of the CO-to-H$_2$ conversion factor. Because the star formation occurs in molecular gas, the total or atomic gas depletion time depend to first order on the balance of atomic and molecular gas in the galaxy combined with the molecular gas depletion time.

\subsection{Abundance and makeup of the ISM}
\label{sec:abundance}

The mass of ISM material in a galaxy also varies. Most notably, the ratio of ISM mass to stellar mass varies as a function of stellar (or halo or total baryonic) mass for galaxies on the star forming main sequence (e.g., Fig. \ref{fig:intscaling}). The sense of this variation is that galaxies with low stellar mass have a higher gas-to-stellar mass ratio. These large gas reservoirs are overwhelmingly \textsc{Hi}, so that a low mass star-forming galaxy can have even more mass in an atomic gas ISM than it has in stars \citep{HUNTER24REVIEW}. 

The makeup of the ISM also varies. In star-forming galaxies the mass of molecular gas tracks the stellar mass fairly well \citep{YOUNG91SFGAS,SAINTONGE22REVIEW}. The molecular to atomic gas ratio correlates with stellar mass for star-forming galaxies. this variation has the sense that a massive galaxy like our Milky Way holds a larger fraction of its overall ISM mass in the molecular phase, while lower mass galaxies have more atomic gas compared to molecular gas. A similar correlation exists with stellar surface density \citep{HUNTER98SFGAS,SAINTONGE22REVIEW}, and with galaxy morphological type in the direction that later type galaxies tend to have lower ratios of molecular to atomic gas \citep{YoungKnezek1989,Obreschkow2009}. The lower molecular-to-atomic gas ratio in lower mass galaxies is often attributed to the effects of lower metallicity and lower pressure. The former enhances photodissociation of molecules and slows down their formation, while the latter (caused by the lower stellar potential) diminishes the density of the gas and the conversion of \textsc{Hi} to H$_2$. 

Many of these trends are, strictly speaking, trends for the luminosity of a low lying CO rotational transition versus the luminosity of the \textsc{Hi} 21~cm transition. While the translation of the latter to atomic gas mass is fairly robust, converting CO luminosity to molecular gas mass is frequently done assuming a constant conversion factor, likely an oversimplification \citep{BOLATTO13REVIEW}. Although the trends mentioned above will not change qualitatively, the values of their slopes depend on assumptions about the CO-to-H$_2$ conversion, which is affected by metallicity, velocity dispersion, temperature of the molecular gas, and possibly other environmental parameters such as X-ray and cosmic ray flux (see \S\ref{sec:h2}).

Metallicity and stellar mass correlate, in the direction of lower stellar mass galaxies being poorer in heavy elements \citep{Tremonti2004,MAIOLINO19REVIEW}. This correlation is attributed mostly to lower mass galaxies being less able to retain nucleosynthetic products, which are preferentially lost to the CGM and IGM via enriched galactic winds that escape the shallower potential wells of these galaxies \citep{TUMLINSON17REVIEW}. The introduction of a third parameter (star formation rate) produces a tighter correlation that shows less evolution with redshift \citep{Mannucci2010}. This is likely caused by the star formation rate acts as a proxy for both the inflow rate of new gas, which is likely relatively pristine and low metallicity and the strength of galactic winds. 

Metallicity and dust-to-gas ratio are also strongly correlated \citep{DRAINE07FITS,Remy-Ruyer2014,galliano2018}. Heavy elements are necessary to make dust grains, and so their abundance affects the gas-to-dust ratio. Though challenging, observations of the dust-to-gas ratio in low metallicity, low mass systems suggest low dust-to-metals ratios. That is, observations suggest that the fraction of heavy elements in the dust phase is lower for low mass, low metallicity galaxies. This can be understood in terms of simple evolutionary models that balance dust production from stars and supernovae, buildup in the ISM, and destruction \citep[e.g.,][]{galliano2018,Aniano2020}. Because observations of this trend are technically challenging and sparse, especially robust observations for low metallicity galaxies, the precise behavior of the dust-to-metals ratio at low metallicity ($12+\log_{10}{\rm O/H}\lesssim8$) remains an important area for future work.

\subsection{Mergers, starburst galaxies, quiescent and quenching galaxies}
\label{sec:mergers}

Galaxies that are located over the main sequence (\S\ref{sec:ks}) display an overabundance of star formation for their stellar mass and are called starbursts. These galaxies are rare, less than 1\% of the population depending on the precise definition. The formation activity of starbursts is such that they use up their molecular gas reservoir very quickly, on a timescale that can be as short as $\sim0.1$~Gyr compared to a typical $1-2$~Gyr for galaxies near the star forming main sequence \citep[][]{TACCONI20REVIEW,SAINTONGE22REVIEW}. Some of the most strongly star-forming outliers are the result of equal mass mergers, where the interaction produces torques that remove angular momentum from the gas driving it down the potential well \citep{SANDERS96REVIEW}. Interactions among dissimilar mass galaxies can also produce starbursts. The prototypical starburst galaxy M82 is an example of this, where the starburst is caused by its interaction with the much larger nearby spiral galaxy M81. Bars also can drive gas into galaxy centers and give rise circumnuclear starburst as discussed above in Section \ref{sec:bars}. 

Starbursts are very gas rich, and are characterized not just by high star formation rates per unit stellar mass but also by high molecular-to-atomic gas ratios, high gas fractions, and short molecular gas depletion times. Much of the star formation activity in starbursts takes place behind large columns of dust. This can make it difficult to study the structure and physical conditions in these systems at optical wavelengths, but also makes these galaxies shine brightly in the infrared. Analysis of the {\it IRAS} far-infrared satellite data found that some of the local most actively star-forming galaxies where unremarkable at optical wavelengths, but displayed IR luminosities $\gtrsim10^{12}$~L$_\odot$. These objects receive the name of ``Ultra-Luminous IR Galaxies'' or ULIRGs \citep{SANDERS96REVIEW}. 

On the other end of the spectrum of activity are quiescent or quenched galaxies. They show low star formation rates for their stellar mass, which places them below the star forming main sequence. These galaxies are usually, but not always, gas poor for both molecular and atomic gas \citep{SAINTONGE22REVIEW}. They also tend to have earlier morphological types, frequently appearing as lenticular or elliptical galaxies. They may display ionized gas emission, but it tends to be low intensity and related to ionization by older stellar populations, hence unrelated to star formation activity \citep{SANCHEZ20REVIEW}. Some of the remnant neutral and molecular gas in these systems may be stabilized against collapse and subsequent star formation by the stellar potential \cite{Martig2009}, so that a galaxy does not need to be completely devoid of cold gas to be quiescent \citep{Colombo2020,SAINTONGE22REVIEW}. Quiescent galaxies are very abundant in high density environments like galaxy clusters, where gas is removed by interactions with other galaxies and the intra-cluster medium.

\section{Conclusions}

Even after almost a century of multi-wavelength study, the ISM in galaxies remains a rich topic of critical importance to the growth of structure in the Universe. Current and plannet facilities including ALMA, JWST, MeerKAT, the VLA, ngVLA and SKA, and any next generation infrared or X-ray probe missions will are powering new advances each year and promise a bright future. The textbooks by \citet{DRAINE11BOOK} and \citet{OSTERBROCK06BOOK} are outstanding resources to dive more into this rich topic, while \citet{Ryden2021}, \citet{CONDON16BOOK}, \citet{Tielens2005}, and the classic book by \citet{SPITZER68BOOK} offer accessible introductory texts.

\begin{ack}[Acknowledgments]{}
This is a pre-print of a chapter for the Encyclopedia of Astrophysics (edited by I. Mandel, section editor S. McGee) to be published by Elsevier as a Reference Module. A.K.L. gratefully acknowledges support by a Humbolt Research Award and grants NSF AST AWD 2205628, JWST-GO-02107.009-A, and JWST-GO-03707.001-A. A.D.B. gratefully acknowledges partial support from grants NSF AST 2108140 and 2307441.
\end{ack}

\newcommand{\physrep}{Physics Reports}
\newcommand{\ssr}{Space Science Reviews}
\newcommand{\apjl}{The Astrophysical Journal Letters}
\newcommand{\apjs}{The Astrophysical Journal Supplement Series}
\newcommand{\apj}{The Astrophysical Journal}
\newcommand{\aj}{The Astronomical Journal}
\newcommand{\aap}{Astronomy \& Astrophysics}
\newcommand{\mnras}{Monthly Notices of the Royal Astronomical Society}
\newcommand{\araa}{Annual Review of Astronomy \& Astrophysics}
\newcommand{\pasp}{Publications of the Astronomical Society of the Pacific}
\newcommand{\pasa}{Publications of the Astronomical Society of Australia}
\newcommand{\pre}{Physical Review Letters}
\newcommand{\prd}{Physical Review D}
\newcommand{\nat}{Nature}
\newcommand\aapr{Astronomy and Astrophysics Reviews}
\newcommand{\jcp}{Journal of Computational Physics}
\newcommand{\jfm}{Journal of Fluid Mechanics}
\newcommand{\rmp}{Reviews of Modern Physics}
\newcommand{\prl}{Physical Review Letters}
\newcommand{\na}{New Astronomy}
\newcommand{\jqsrt}{Journal of Quantitative Spectroscopy and Radiative Transfer}

\bibliographystyle{Harvard}
\input{els_article.bbl}

\end{document}